\documentclass[12pt]{article}
\usepackage{amscd}
\usepackage{verbatim}
\usepackage{amssymb}
\usepackage{amsmath}
\usepackage{tabularx}
\usepackage[utf8]{inputenc}
\usepackage[a4paper, total={6in, 8in}]{geometry}
\usepackage{graphicx}
\usepackage{amsthm}
\usepackage{tensor}
\usepackage[utf8]{inputenc}
\usepackage{appendix}
\usepackage{hyperref}
\usepackage[parfill]{parskip}

\begin{document}
\begin{center}
\vspace{24pt} { \large \bf The Black hole Black string phase transition in Einstein-Gauss-Bonnet gravity} \\
\vspace{30pt}
\vspace{30pt}
\vspace{30pt}
{\bf Sreejith Nair \footnote{sreejith.nair@students.iiserpune.ac.in}}, {\bf Vardarajan
Suneeta\footnote{suneeta@iiserpune.ac.in}}\\
\vspace{24pt} %%
{\em  The Indian Institute of Science Education and Research (IISER),\\
Pune, India - 411008.}
\end{center}
\date{\today}
\bigskip
\begin{center}
{\bf Abstract}
\end{center}
We investigate the presence of a black hole black string phase transition in Einstein Gauss Bonnet (EGB) gravity in the large dimension limit. The merger point is the static spacetime connecting the black string phase with the black hole phase. We consider several ranges of the Gauss-Bonnet parameter. We find that there is a range when the Gauss-Bonnet corrections are subordinate to the Einstein gravity terms in the large dimension limit, and yet the merger point geometry does not approach a black hole away from the neck. We cannot rule out a topology changing phase transition as argued by Kol. However as the merger point geometry does not approach the black hole geometry asymptotically it is not obvious that the transition is directly to a black hole phase. We also demonstrate that for another range of the Gauss-Bonnet parameter, the merger point geometry approaches the black hole geometry asymptotically when a certain parameter depending on the Gauss-Bonnet parameter $\alpha$ and on the parameters in the Einstein-Gauss-Bonnet black hole metric is small enough.
\newpage

\tableofcontents

\section{Introduction}
\label{section:intro}
 \hspace{1cm} The discovery of a long-wavelength instability in uniform black strings and flat p-branes by Gregory and Laflamme\cite{grins} has led to a huge body of work with the aim of finding the endpoint of the instability. There is a static non-uniform perturbation - a threshold mode such that the string is unstable for wavelengths more than this critical wavelength. This suggests that there could be a static non-uniform black string, which has a horizon that is non-uniform in the extra compact dimension along the string.
 The expectation is that one could increase the non-uniformity parameter to generate new non-uniform black string solutions. These solutions emanate from the branch of static uniform black strings solutions in Einstein Gravity. It was proposed that as one increases the non-uniformity, eventually the horizon pinches off, and the black string transitions into a black hole. This is a phase transition in the space of static solutions to Einstein gravity on geometries with a horizon which also asymptote to Minkowski spacetime times a circle at infinity. The transition involves a topology change in the horizon of the spacetime, and the intermediate spacetime between these two phases is known as the merger point\cite{phasetranseinpap}-\cite{newphase}. A review of these developments can be found in \cite{phaseriv}, \cite{phasetransein}. These arguments were backed up by several numerical studies on the space of static solutions to Einstein Gravity\cite{num1}-\cite{num3}.
Kol \cite{phasetranseinpap}\cite{phasetransein} studied this horizon topology change in the Euclidean version of the spacetime. He argued that the merger point in the near neck region is a conical geometry. Emparan and Suzuki successfully demonstrated the existence of such a merger point in Einstein Gravity, in the large dimension limit \cite{Ricciflow}. The large dimension limit of Einstein gravity was developed as an analytical tool by Emparan et al.\cite{largedfirst}-\cite{largedrev} to study various aspects of black holes.  The black hole-black string transition was shown to be mediated through the Ricci flow equation \cite{Ricciflow} (studied in the mathematics literature) in the large dimension limit; this allowed for well-known solutions to Ricci flow in two dimensions to be used in the study of the black hole-black string phase transition.

 \hspace{1cm} Our paper deals with this phase space of static solutions. We will not be addressing the issue of the dynamical endpoint of the Gregory-Laflamme instability. However, we summarize the interesting work in this direction in Einstein gravity. An attempt was made to guess the potential end state by using entropic arguments and studying the space of all possible static solutions in Einstein gravity\cite{phasestructure}-\cite{phaseriv}. The dynamical
evolution to the end state is expected and seen numerically to happen through a pinching off of the horizon \cite{phasetranseinpap}\cite{phasetransein}. Numerical studies have demonstrated a fractal structure for the horizon with curvature which blows  up\cite{blkstrifate}\cite{selfsimcosvil}. The above mentioned behaviour under dynamical evolution is very similar to the Plateau–Rayleigh instability in fluid dynamics\cite{selfsimcosvil}. Emparan, Suzuki and Tanabe have also analytically demonstrated the expected behaviour in their work using the large dimension limit of general relativity \cite{largdpert}.

 \hspace{1cm}
We focus in this paper on the static merger point in the phase space of solutions, not in Einstein gravity but in Einstein-Gauss-Bonnet (EGB) gravity. EGB gravity appears naturally for example, in string theory as higher curvature corrections to Einstein gravity \cite{stringcor}\cite{blackholesinegbgravity}. Since the merger point has a conical singularity, it is important to consider higher curvature corrections. Recall that in EGB gravity the Lagrangian is:
\begin{equation}
     L_{EGB}=R+\alpha(R^2-4R^{\mu\nu}R_{\mu\nu}+R^{\mu\nu\rho\sigma}R_{\mu\nu\rho\sigma})
     \label{egblag}
 \end{equation}
where $\alpha$ is the EGB parameter.

 \hspace{1cm}Black strings and branes in EGB gravity were discussed in \cite{troncoso}. Recently, the instability of uniform black strings was also seen in EGB gravity
 \cite{formofegbblackhole}-\cite{GLinGBlarged}. The presence of a phase of non-uniform black strings is also discussed in \cite{formofegbblackhole}.
  Upon increasing the non-uniformity in the non-uniform black string emanating at the critical wavelength, will this eventually lead to a topology changing phase transition to a black hole? This is the question we address in this paper.
  We would also like to add that generalized black string solutions have been found in other theories of gravity like scalar-tensor gravity, see for example \cite{nakas1}, \cite{nakas2} and \cite{nakas3}.

  \hspace{1cm}We use the methodology developed by Emparan and Suzuki in their work \cite{Ricciflow} and consider EGB corrections to their results. They showed that in a neck of radial extent $r_h/\sqrt{n}$ and  $r_h/n$ along the extra dimension, one can have a local conical geometry \cite{phasetranseinpap}\cite{phasetransein}, which can be correctly extended to the geometry of a large black hole on a cylinder. Here $r_h$ is the horizon radius of the black hole and $n$ is the (large) dimension.  The problem of finding the merger point geometry reduces to finding solutions to the well-known logarithmic diffusion equation obtained from the Ricci flow. Emparan and Suzuki demonstrated that the King-Rosenau solution\cite{king1}-\cite{king3} to the logarithmic diffusion equation has the desired behaviour of a merger point. This solution produces the near horizon limit of a black hole asymptotically; and a local conical geometry at the neck. Thus it suggests a topology changing phase transition, and further that the transition is to a black hole. This shows the existence of a merger point in the large dimension limit of Einstein gravity.

 \hspace{1cm}We extend the analysis in \cite{Ricciflow} to EGB gravity. We obtain a modification to the flow equation\cite{Ricciflow} for various ranges of $\alpha$. We demonstrate that the merger point has the expected behaviour for $\alpha\sim\mathcal{O}(1/n^2)$ or smaller, where $n$ is the dimension and is taken to be large. More precisely, this happens when a parameter constructed from $\alpha$ and $r_0$, namely $\epsilon = \alpha n^2/r_{h}^{2}$ is small.  At $\alpha > \mathcal{O}(1/n^2)$, we obtain a "modified" logarithmic diffusion equation in the place of logarithmic diffusion equation. This equation does not seem to allow for an obvious phase transition to the black hole phase. We cannot rule out a topology changing phase transition, but it is not obvious that the transition is to a black hole geometry.
We demonstrate this by showing that the near neck solutions' asymptotic behaviour will not match the near horizon limit of a black hole when $\alpha$ is large enough. The analysis proceeds in the following steps:

\begin{itemize}
    \item First(section.\ref{sec:nearhor}), we will study the near horizon behaviour of a black hole in EGB gravity at various orders of $\alpha$.
    \item Second(section \ref{subsec:derifloeq}\&\ref{subsec:floeqbhbs}), we will work in an $N+1$ formulation of EGB gravity. We will derive the evolution equation for extrinsic curvature in EGB gravity for various orders of $\alpha$ and show that there is a modification to the flow equation obtained in Einstein gravity (the Ricci flow) for the range $\mathcal{O}(1/n)>\alpha>\mathcal{O}(1/n^2)$, even when the corrections from the Gauss-Bonnet terms are subordinate to the Einstein term. For $\alpha\sim \mathcal{O}(1/n^2)$, there is no modification to the Ricci flow at leading order in $1/n$.
     \item Third(section \ref{subsec:pertlog}\&\ref{subsec:ordless}), we will demonstrate that a solution of the logarithmic diffusion equation can be perturbed to obtain another solution, which, for $\alpha$ in an appropriate range matches approximately with the near-horizon limit of the EGB black hole when $\epsilon$ is small enough.
    \item Fourth(section \ref{subsection:ordgat}), we will demonstrate for other ranges of $\alpha$ that the solution to the modified flow equation cannot match to the near horizon limit of the EGB black hole. In these cases, one could still have a topology changing phase transition, but it is not obvious that the transition is from a black string to the black hole.

    \item In section \ref{cone} we discuss the near-neck geometry in EGB gravity.
\end{itemize}

  \section{Near Horizon limit of EGB black hole}
  \label{sec:nearhor}
\hspace{1cm}
In this section, we look at the near-horizon limit of the black hole that was found in EGB gravity \cite{blackholesinegbgravity}. We focus on the asymptotically flat solution with $\alpha > 0$  (as this is the one that arises in string theory) the 'minus branch'. This is
 \begin{equation}
    ds^2=-f(r)dt^2+\frac{dr^2}{f(r)}+r^2d\Omega^2_{d-2}
    \label{blackhole}
 \end{equation}
 with $f(r)$ being of the following form\cite{blackholesinegbgravity}\cite{formofegbblackhole};

\begin{equation}
    f(r)=1+\frac{2r^2}{\alpha(d-3)(d-4)}\Biggl(1-\sqrt{1+\frac{\alpha}{r_h^2}(d-3)(d-4)\Big(1+\frac{\alpha}{4r_h^2}(d-3)(d-4)\Big)\Big(\frac{r_h}{r}\Big)^{d-1}}\Biggl)
    \label{formoff}
\end{equation}

  Let us attempt to find the near horizon limit of this metric in the large $n$ limit, where $n=d-2$ by going to the coordinates where $R = (\frac{r}{r_h})^n$, Now for different orders of $\alpha$ we get different results in the large $n$ limit, if $\alpha\sim \mathcal{O}(1/n^2)$ we have;

  \begin{equation}
  f(R)=\frac{(R-1)}{R}\Big(1-\frac{\alpha n^2}{4r_h^2R}\Big) +.....
    \label{formoffo2}
 \end{equation}
 Here we performed a binomial expansion of the expression inside the square root and that requires
 $\frac{\alpha n^2}{4r_h^2}$ small.
  Now the metric will take the following form;
   \begin{equation}
    ds^2=-f(R)dt^2+\frac{r_h^2R^{2\frac{(1-n)}{n}}dR^2}{n^2f(R)}+r_h^2R^{\frac{2}{n}}d\Omega^2_{n}
    \label{bh}
 \end{equation}

 Next, going to the near horizon limit \cite{largeDgravstring}\cite{largedrev} by setting $ln(R)<<n$, followed by $R=cosh^2(x/2), t=\frac{\tilde t}{r_h}$, we get the following;

  \begin{equation}
    ds^2=r_h^2\Biggl(-tanh^2(x/2)\Big(1-\frac{\alpha n^2}{4r_h^2cosh^2(x/2)}\Big)d\Tilde{t}^2+\frac{dx^2}{n^2}+d\Omega^2_{d-2}\Biggl)
    \label{blackholenear}
 \end{equation}
   when $\frac{\alpha n^2}{4r_h^2}$ is small and we assume its higher powers and products with powers of $1/n$ are negligible. But,
 if $\alpha>\mathcal{O}(1/n^2)$ we get;
 \begin{equation}
     f(R)=\frac{R^{1/2}-1}{R^{1/2}}
     \label{formoff1}
 \end{equation}
  \hspace{1cm}The metric (\ref{blackholenear}) reduces to the Einstein gravity Schwarzschild-Tangherlini black hole in the limit of $\alpha \longrightarrow 0$ but (\ref{formoff1}) produces a metric that does not have the Einstein gravity black hole result as a limiting case. This means that the black hole with  $\alpha>\mathcal{O}(1/n^2)$ has a near horizon behaviour which is very different from the near horizon behaviour in Einstein gravity. Note that the constant $\alpha$ disappears from the leading near-horizon behaviour in (\ref{formoff1}).  This observation will be significant later.

 \section{The EGB gravity equations}
 \label{sec:floeq}
 \hspace{1cm}Before going ahead with the calculations, let us briefly discuss the work that was done in Einstein Gravity by Emparan and Suzuki\cite{Ricciflow}. They explored the geometry of the merger point in the black hole black string phase transition, which was conjectured by Kol to be locally a double cone geometry near the neck \cite{phasetranseinpap}\cite{phasetransein}. They considered a metric of the form\footnote{It should be noted that the $n$ in this metric is different from the $n$ used in the large $d$ limit of (\ref{blackhole}), but both converge in the large $d$ limit.};
 \begin{equation}
        ds^2=N^2(\rho,y)d\rho^2+\frac{1}{n}g_{ab}(\rho,y)dy^ady^b+H^2(\rho,y)e^{2C(\rho,y)/n}d\Omega^2_n
        \label{formnearmerg}
\end{equation}
 which is a generalization in the large dimension limit of the double cone geometry proposed by Kol. Then they showed that this form of the metric is valid locally in a small neighbourhood near the neck of the non-uniform string with an extension of $r_h/n$ along the string direction and  $r_h/\sqrt{n}$ in the radial direction. For a metric of the form (\ref{formnearmerg}), they studied the Einstein equations in a Hamiltonian formulation. At leading order in $n$ in the large $n$ limit, it was shown that the solutions obey the Ricci flow equation, which is well-studied by mathematicians. This equation has as a solution, the King-Rosenau geometry \cite{king1}-\cite{king3}. Emparan and Suzuki showed that this geometry in a limit reduces to a smoothed cone on either side of the merger point. At the merger point, Emparan and Suzuki showed that the double cone geometry conjectured by Kol is recovered locally in the neighbourhood of the neck.

 \hspace{1cm} The form of the metric (\ref{formnearmerg}) and the associated scaling of metric coefficients with $n$ in Einstein gravity was motivated by the double cone geometry. It is reasonable to expect the same behaviour (with GB corrections) as long as the GB terms are subordinate to the Einstein terms. This means that the ansatz and associated methodology can be adapted to EGB gravity as long as the GB corrections are subordinate to the Einstein terms. However, for completeness, we will be presenting the flow equations for all ranges of $\alpha$.

 \subsection{N+1 formulation of EGB Gravity}
 \label{subsec:derifloeq}
 \hspace{1cm}Let us begin by considering the ansatz of the form (\ref{formnearmerg}). Now we use the N+1 formulation of EGB gravity\cite{hamiltonian} for the metric (\ref{formnearmerg}) and study its evolution with $\rho$ according to the EGB equations. Now if we foliate the spacetime along the $\rho$ direction, the hypersurface has the following extrinsic curvatures and its trace\footnote{Please notice that the convention we are using is not the same as \cite{Ricciflow}, instead we have adopted the standard convention from \cite{hamiltonian}};

 \begin{equation}
     K_{AB}=-\frac{1}{2N}\partial_0\gamma_{AB};\hspace{0.25cm}K=K^A{}_A
 \end{equation}

 \hspace{1cm}Here, {A,B,C,...} are coordinates that run over the entire foliated hypersurface and $\gamma_{AB}$ are the components of the metric restricted to the hypersurface.For this type of a foliation we have, in the large $n$ limit, $K_{ab}\sim O(1/n) \& K_{ij}\sim O(1)$ which would make $K\sim O(n)$. Now upon making it satisfy the vacuum scalar and vector constraints of EGB gravity\cite{hamiltonian};

\begin{equation}
        M+\alpha(M^2-4M_{AB}M^{AB}+M_{ABCD}M^{ABCD})=0
        \label{scalarcons}
\end{equation}
\begin{equation}
       N_A+2\alpha(MN_A-2\tensor{M}{_A^B}N_B+2M^{CD}N_{ACD}-\tensor{M}{_A^D^B^C}N_{BCD})=0
       \label{vectorcons}
\end{equation}
such that;
\begin{equation}
        M_{ABCD}=R_{ABCD}-K_{AC}K_{BD}+K_{AD}K_{BC}
        \label{MABCD}
    \end{equation}
    \begin{equation}
        M_{AB}=\gamma^{CD}M_{ACBD}=R_{AB}-KK_{AB}+K_{AC}\tensor{K}{^C_B}
        \label{MAB}
    \end{equation}
    \begin{equation}
        M=\gamma^{CD}M_{CD}=R-K^2+K_{CD}K^{CD}
        \label{M}
    \end{equation}
    \begin{equation}
        N_{ABC}=D_AK_{BC}-D_BK_{AC}
        \label{NABC}
    \end{equation}
    \begin{equation}
        N_A=\gamma^{CD}N_{CAD}=D_B\tensor{K}{^B_A}-D_AK
        \label{NA}
    \end{equation}

 Next, we shall attempt to gauge fix the metric (\ref{formnearmerg}), up to leading order in $n$ .

\hspace{1cm} For the metric (\ref{formnearmerg}), $M$ will contain an $O(n^3)$ term of the form $(\nabla_a H)^2$. It can be seen that for all orders of $\alpha$, this term will appear at the leading order in the scalar constraint. Upon setting this term to zero, we have $H$ to be independent of $y$. The next term in $M$ is of the form, $\frac{H'^2}{N^2}-1$(this term is at $O(n^2)$). For all $\alpha$ this will then be the leading order term in the scalar constraint, therefore (\ref{scalarcons}) will imply;

\begin{equation}
    \label{scalcon}
    \frac{H'^2}{N^2}-1=0
\end{equation}

which can be solved by
\begin{equation}
    H=\rho,\hspace{0.5cm}
    N=1+\frac{N_1(\rho,y)}{n} .
\end{equation}
We note that we have introduced the subleading term $N_1$, as it is required for consistency with the evolution equation for $K_{ij}$. The $K_{ij}$ flow equation will give a differential equation for $N_1$ in terms of the $g_{ab}$ and its derivatives.

Now we can further do the coordinate transformations;

\begin{equation}
    \rho\longrightarrow\rho\Big(1+\frac{\beta(\rho,y)}{n}\Big),\hspace{0.2cm}y^a\longrightarrow y^a+\zeta^a(\rho,y).
\end{equation}

to set $C(\rho,y)=0$. This coordinate transformation is identical to the one used in Einstein gravity\cite{Ricciflow} and we assume $\zeta^a$ has already been chosen appropriately to cancel out potential $g_{\rho a}$ terms. Now the metric takes the form\footnote{It can also be checked that this form of the metric satisfies all the vector constraints up to leading order.};

\begin{equation}
        ds^2=(1+\frac{2N_1(\rho,y)}{n})d\rho^2+\frac{1}{n}g_{ab}(\rho,y)dy^ady^b+\rho^2d\Omega^2_n
        \label{formnearmergred}
\end{equation}

 Thus the above gauge fixed metric is identical to the one obtained in Einstein Gravity\cite{Ricciflow}. Next, we shall evaluate the evolution equation for $K_{AB}$ of the gauge fixed metric and see what we obtain. The evolution equation for $K_{AB}$ reads \cite{hamiltonian};
\begin{eqnarray}
        \label{hameq}
            M_{AB}-\frac{1}{2}M\gamma_{AB}+K_{AC}\tensor{K}{^C_B}-\gamma_{AB}K_{CD}K^{CD}\nonumber\\+L_nK_{AB}
            -\gamma_{AB}\gamma^{CD}L_nK_{CD}+2\alpha\Big[H_{AB}+ML_nK_{AB}\nonumber\\
            -2\tensor{M}{_A^C}L_nK_{CB}
            -2\tensor{M}{_B^C}L_nK_{CA}-\tensor{W}{_A_B^C^D}L_nK_{CD}\Big]&=&0
    \end{eqnarray}

 where $H_{AB} \& \tensor{W}{_A_B^C^D}$ are defined as follows;

    \begin{eqnarray}
        H_{AB}=MM_{AB}-2(M_{AC}\tensor{M}{^C_B}+M^{CD}M_{ACBD})+M_{ACDE}\tensor{M}{_B^C^D^E}\nonumber\\
        -2\Big[-K_{CD}K^{CD}M_{AB}-\frac{1}{2}MK_{AC}\tensor{K}{^C_B}+K_{AC}\tensor{K}{^C_D} \tensor{M}{^D_B}+K_{BC}\tensor{K}{^C_D}\tensor{M}{^D_A}\nonumber\\
        +K^{DC}\tensor{K}{_C^E}M_{ADBE}+N_AN_B
        -N^C(N_{CAB}+N_{CBA})-\frac{1}{2}N_{CDA}\tensor{N}{^C^D_B}\nonumber\\
        -N_{ACD}\tensor{N}{_B^C^D}\Big]
        -\frac{1}{4}\gamma_{AB}\Big[M^2-4M_{CD}M^{CD}
        +M_{CDEF}M^{CDEF}\Big]\nonumber\\
        -\gamma_{AB}\Big[K_{CD}K^{CD}M-2M_{CD}K^{CE}\tensor{K}{_E^D}-2N_CN^C+N_{CDE}N^{CDE}\Big]
    \end{eqnarray}

 \begin{equation}
       \tensor{W}{_A_B^C^D} =M\gamma_{AB}\gamma^{CD}-2M_{AB}\gamma^{CD}-2\gamma_{AB}M^{CD}+2M_{AEBF}\gamma^{EC}\gamma^{FD}
        \label{WABCD}
    \end{equation}

    Now upon subtracting the trace of (\ref{hameq}) from half of the scalar constraint (\ref{scalarcons}) we get the following equation;

  \begin{eqnarray}
      (d-2)\Big[\frac{1}{2} M+K_{AB} K^{AB}+\gamma^{AB} L_{n} K_{AB}\Big]+2(d-4) \alpha\Big[\frac{1}{4}(M^{2}-4 M_{AB} M^{AB}\nonumber\\
      +M_{ABCD} M^{ABCD})
+M K_{AB} K^{AB}-2 K_{A}{}^{B} \tensor{K}{^A_C} M^{C}{ }_{B}-2 N_{A} N^{A}\nonumber\\
+N_{ABC} N^{ABC}
+M \gamma^{AB} L_{n} K_{AB}-2 M^{AB} L_{n} K_{AB}\Big]=0
  \end{eqnarray}

    Note that the above equation in the large d limit can be written in the following form;

    \begin{eqnarray}
     \begin{alignedat}{2}
        K_{AB}K^{AB}+\gamma^{AB}L_nK_{AB}
        +2\alpha\Big[MK_{AB}K^{AB}\\-2M_{AB}K^{AC}\tensor{K}{_C^B}
        -2N_AN^A+N_{ABC}N^{ABC}\Big]\\
        &=&2\alpha\Big[(2M^{AB}L_nK_{AB}
      -M\gamma^{AB}L_nK_{AB}\Big]
    \end{alignedat}
    \end{eqnarray}

     Using the above equation and the scalar constraint(\ref{scalarcons}) on the evolution equation for the extrinsic curvature (\ref{hameq}), we can reduce the evolution equation to the following form;

    \begin{eqnarray}
        \label{redfloweq}
        M_{AB}-(-K_{AC}\tensor{K}{^C_B}-L_nK_{AB})+2\alpha\Big[\Acute{H}_{AB}+ML_nK_{AB}\nonumber\\
        -2\tensor{M}{_A^C}L_nK_{CB}-2\tensor{M}{_B^C}L_nK_{CA}-\tensor{W}{_A_B^C^D}L_nK_{CD}\Big]&=&0
    \end{eqnarray}

    where;
    \begin{eqnarray}
        \Acute{H}_{AB} = MM_{AB}-2(M_{AC}\tensor{M}{^C_B}+M^{CD}M_{ACBD})+M_{ACDE}\tensor{M}{_B^C^D^E}
        \nonumber\\
        -2\Big[-K_{CD}K^{CD}M_{AB}
        -\frac{1}{2}MK_{AC}\tensor{K}{^C_B}+K_{AC}\tensor{K}{^C_D} \tensor{M}{^D_B}\nonumber\\
        +K_{BC}\tensor{K}{^C_D}\tensor{M}{^D_A}
        +K^{DC}\tensor{K}{_C^E}M_{ADBE}\nonumber\\
        +N_AN_B-N^C(N_{CAB}+N_{CBA})
        -\frac{1}{2}N_{CDA}\tensor{N}{^C^D_B}\nonumber\\
        -N_{ACD}\tensor{N}{_B^C^D}\Big]
        -\gamma_{AB}(2M^{CD}L_nK_{CD}-M\gamma^{CD}L_nK_{CD})
        \label{HAB}
    \end{eqnarray}

      Now, for the gauge fixed metric(\ref{formnearmergred}) we have the following (at leading order in $n$ in the large $n$ limit);

   \begin{eqnarray*}
        K_{ab}=-\frac{\partial_\rho g_{ab}}{2n};\hspace{0.5cm}K_{ij}=-\rho g_{ij}(1-\frac{N_1}{n});\hspace{0.5cm}K=-\frac{n}{\rho}\big(1-\frac{1}{n}(N_1-\frac{\rho\partial_\rho(ln g)}{2})\big)\nonumber\\
    \end{eqnarray*}
    \begin{eqnarray*}
        M_{ab}=R_{ab}-\frac{\partial_\rho g_{ab}}{2\rho};\hspace{0.5cm}
         M_{ij}=(2N_1-\frac{\rho\partial_\rho ln g}{2})g_{ij};\hspace{0.5cm}M=n(g^{ab}R_{ab}  -\frac{\partial_\rho ln g}{\rho}+\frac{2N_1}{\rho^2})\nonumber\\
    \end{eqnarray*}
    \begin{eqnarray*}
         N_{aij}=\rho\frac{\partial_aN_1}{n}g_{ij};N_{abc}=-\frac{1}{2n}(D_a\partial_\rho g_{bc}-D_b\partial_\rho g_{ac});
         N_a=-\frac{g^{cd}}{2}(D_c\partial_\rho g_{ad}-D_a\partial_\rho g_{cd})-\frac{\partial_aN_1}{\rho}
    \end{eqnarray*}
    \begin{eqnarray*}
    M_{abcd}=\frac{R_{abcd}}{n};\hspace{0.5cm}M_{ijkl}=(1-\rho^2)(g_{ik}g_{jl}-g_{il}g_{jk});\hspace{0.5cm}M_{aibj}=-\frac{\rho\partial_\rho g_{ab}}{2n}g_{ij}
    \end{eqnarray*}

    Now the leading order analysis of the evolution equation(\ref{redfloweq}) using the  gauge fixed metric (\ref{formnearmergred}) will result in the following for various orders of $\alpha$ (Table.\ref{tabfloweq}), where we have the following definitions($\dot{A}=\partial_\rho A$);

     \begin{equation}
    \Acute{M}_1=g^{ab}R_{ab}- \frac{\partial_\rho ln(g)}{\rho}+\frac{2N_1}{\rho^2}
    \end{equation}
    \begin{equation}
    \mathcal{R}_{ab}=R_{ac}\tensor{R}{^c_b}+R^{cd}R_{acbd}-\frac{1}{2}R_{acde}\tensor{R}{_b^c^d^e}
    \end{equation}
    \begin{equation}
    \mathcal{G}_{ab}=\frac{\partial_\rho g^{cd}}{2\rho}R_{acbd}-\frac{1}{2\rho}(\tensor{R}{_a^c}\partial_\rho g_{cb}+\tensor{R}{_b^c}\partial_\rho g_{ca})
    \end{equation}
   % \begin{equation}
    %    \phi=N_1-\frac{\rho^2}{2}R^{ab}g_{ab}
    %\end{equation}
  %     \Phi=\big(\acute{M}_1-2\frac{(1-\rho^2)}{\rho^4}+\frac{4}{\rho^2}\big)\big(2N_1-\rho\frac{\partial ln(g)}{2}\big)\\+\rho \dot{g}_{ab}(R^{ab}+\frac{\dot{g}^{ab}}{2\rho})+\acute{M}_1-2\frac{(1-\rho^2)}{\rho^4}\\
  %      +\rho^2\big(-\frac{\ddot{g}_{ab}}{2}+D_aD_bN_1\big)\big(\acute{M}_1g^{ab}-2R^{ab}-\frac{\dot{g}^{ab}}{\rho}\big)-\acute{M}_1(2N_1+\rho\dot{N}_1)
  %  \end{multline}
    \begin{table}[h!]
        \centering
        \begin{tabular}{c|c}

        Order of $\alpha$ & Evolution of $K_{ab}$ \\% & Evolution of $K_{ij}$\\
        &\\

      $\alpha\sim\mathcal{O}(1)$   &  $\frac{1}{2\rho^2} \dot{g}_{ac}\dot{g}_{bd}g^{cd}=\acute{M}_1(R_{ab}-\frac{\dot{g}_{ab}}{2\rho})-2(\mathcal{R}_{ab}+\mathcal{G}_{ab})$\\ %& $\Phi=0$\\
         &\\

      $\mathcal{O}(1)>\alpha\geq\mathcal{O}(1/n)$  &  $\frac{1}{2n\alpha}(R_{ab}-\frac{\dot{g}_{ab}}{2\rho})+\Big(\acute{M}_1(R_{ab}-\frac{\dot{g}_{ab}}{2\rho})-2(\mathcal{R}_{ab}+\mathcal{G}_{ab})-\frac{1}{2\rho^2} \dot{g}_{ac}\dot{g}_{bd}g^{cd}\Big)=0$ \\%& $\frac{1}{2n\alpha}\phi+\Phi=0$\\
        &\\

      $\mathcal{O}(1/n)>\alpha>\mathcal{O}(1/n^2)$   & $R_{ab}-\frac{\dot{g}_{ab}}{2\rho}+2n\alpha\Big(\acute{M}_1(R_{ab}-\frac{\dot{g}_{ab}}{2\rho})-2(\mathcal{R}_{ab}+\mathcal{G}_{ab})-\frac{1}{2\rho^2} \dot{g}_{ac}\dot{g}_{bd}g^{cd}\Big)=0$\\%& $\phi+2n\alpha\Phi=0$\\
         &\\

     $\alpha\sim \mathcal{O}(1/n^2)$ & $R_{ab}-\frac{\dot{g}_{ab}}{2\rho}=0$\\ % & $\phi=0$\\
     &\\
      \end{tabular}
      \caption{Table of evolution equations}
      \label{tabfloweq}
    \end{table}

    \hspace{1cm}It should also be noted that at $\alpha\geq \mathcal{O}(1/n)$, the Gauss-Bonnet terms either dominate the Einstein gravity terms or are comparable to them --- this is not a physically significant regime. Further, as discussed earlier in this range of $\alpha$, we do not know if the ansatz (\ref{formnearmerg}) is valid. We want the Gauss-Bonnet terms to only provide corrections to the Einstein gravity result. So the relevant solutions are only at orders of $\alpha< \mathcal{O}(1/n)$. In this regime, for the range $\mathcal{O}(1/n)>\alpha>\mathcal{O}(1/n^2)$ the Gauss-Bonnet corrections are significant and change the flow equation from Ricci flow to the more complicated evolution equation listed in the table\footnote{For $\alpha< O(1/n)$ (the physically significant regime), the sub-leading corrections to the flow equation come at $O(1/n)$. Similar corrections are present at all the orders of $\alpha$ in the table. These are not considered in the large $n$ limit.}.

    \hspace{1cm}We have not explicitly written down the evolution equation for $K_{ij}$; this is because it just determines $N_1$ (which is a subleading correction) in terms of $g_{ab}$ and its derivatives. Actually, the $K_{ab}$ evolution equation is coupled with the $K_{ij}$ evolution equation through the $N_1$ in $\Acute{M}_1$. But luckily this will not be of concern to the analysis we are performing, as $\Acute{M}_1$ multiplies $(R_{ab}-\frac{\dot{g}_{ab}}{2\rho})$. This particular form of the coupling makes the system manageable; else, we would have had two very complicated coupled partial differential equations. The coupling of evolution equations is distinct from Einstein gravity and is a feature of EGB gravity\cite{hamiltonian}.

    \subsection{The Black hole-Black string transition}
    \label{subsec:floeqbhbs}
    \hspace{1cm}If we study the black hole-black string transition, we have the $\{a,b,c...\}$ subspace to be two- dimensional. This means that we can work in coordinates where the metric is conformal to the flat metric;

    \begin{equation}
    \frac{1}{n}g_{ab}dx^adx^b=\frac{1}{n}V(\rho,z)(-dt^2+dz^2)
    \label{conformal}
    \end{equation}
   Above, $V$ is independent of $t$ as we are working with a static spacetime\footnote{Please note that we might have to rescale these coordinates appropriately to match with the rest of the geometry away from the region where \ref{formnearmerg} is valid.}. Now if $\alpha\sim \mathcal{O}(1/n^2)$, we get the Einstein gravity result \cite{Ricciflow};

    \begin{equation}
    \partial_\lambda V=\partial^2_z(ln(V))
    \label{log}
    \end{equation}
     while for $\mathcal{O}(1/n)>\alpha>\mathcal{O}(1/n^2)$  from table.\ref{tabfloweq} we get a 'modified' logarithmic diffusion equation;

    \begin{equation}
    \partial_\lambda V=\partial^2_z(ln(V))+2n\alpha\Biggl[-\acute{M}_1(\partial_\lambda V-\partial^2_z(ln(V)))+\frac{[(\partial_\lambda V)^2-3\partial_\lambda V\partial^2_z(ln(V))+(\partial^2_z(ln(V)))^2]}{V}\Biggl]
    \label{modlog}
    \end{equation}
       where, $\lambda=\lambda_0-\frac{\rho^2}{2}$.

    \hspace{1cm}Now let us consider a solution to the logarithmic diffusion equation, which is also a solution to the 'modified' logarithmic diffusion equation. Let us call this solution $V'$. Now using the (\ref{log}) we can reduce (\ref{modlog}) to imply the following equation;

    \begin{equation}
        \partial_{\lambda}V'=\partial^2_{z}(ln(V'))=0
        \label{modlogprob}
    \end{equation}

    \hspace{1cm}Observe that the above equation implies that $V'$ is not dependent on $\rho$. But as we increase $\rho$, we expect to move away from the local geometry near the neck and transition to the black hole geometry.

    \hspace{1cm} If $V$ is a solution to (\ref{modlog}), then it cannot independently solve the logarithmic diffusion part and the modification(part of (\ref{modlog}) inside the square brackets), unless it is independent of $\rho$ and $\alpha$. This observation will come in handy later. The other solutions which have $V$ solving the \textbf{modified} log diffusion equation (\ref{modlog}) but not the logarithmic diffusion equation have to be of the form
    $V(\rho, z, \alpha)$ --- i.e., the solutions must depend on $\alpha$.

    \subsection{Perturbed solutions of the logarithmic diffusion equation}
    \label{subsec:pertlog}

    \hspace{1cm}The relevant flow equation in Einstein gravity is the logarithmic diffusion equation, and we regain the same flow equation in EGB gravity for a sufficiently small GB parameter. It is of interest to study the perturbative corrections to the solutions in Einstein gravity as we expect corrections from GB terms. Now let us study the solution with some small correction, $V=V_0(\rho,z)-\epsilon F(\rho,z)$. Let us treat $\epsilon$ to be some small parameter, and we work in linearized perturbation theory with perturbation parameter $\epsilon$.

    \hspace{1cm}By demanding that the Einstein gravity flow equation be satisfied by $V_0$ and $V$, at linear order in $\epsilon$, we get an equation, which reads;
    \begin{equation}
        \partial_\lambda F=\nabla^2\Big(\frac{F}{V_0}\Big)
        \label{condpert}
    \end{equation}

    Now for the near horizon single black hole solution in Einstein gravity \cite{Ricciflow};

    \begin{equation}
        V^{bh}_0(\rho,z)=\frac{1}{1+e^{-2(\rho^2+z)}}
        \label{nearhorsingblackho}
    \end{equation}
    It can be checked that this gives a corresponding solution, $F^{bh}$ as;

    \begin{equation}
        F^{bh}(\rho,z)=\frac{1}{(1+e^{-2(\rho^2+z)})(1+e^{2(\rho^2+z)})}
    \end{equation}

   This means that we have (for a small parameter $\epsilon$) a solution of the form;

    \begin{equation}
        V^{bh}(\rho,z)=\frac{1}{1+e^{-2(\rho^2+z)}}\Big(1-\frac{\epsilon}{1+e^{2(\rho^2+z)}}\Big)
        \label{modifromegb}
    \end{equation}

    In Section \ref{cone} we consider perturbative corrections to the $V^{cone}=(\rho^2-\mu)/cosh^2(z)$, which is the near neck geometry in Einstein gravity\cite{Ricciflow}.

    \section{Near horizon single Black hole solution in EGB gravity}
    \label{sec:nearhorbhegb}

     \hspace{1cm}We can now attempt to study the conformal factor $V^{bh}$ derived in the previous section. It was demonstrated in \cite{Ricciflow} that the metric with conformal factor (\ref{nearhorsingblackho}) $V_0$ reduces to the near-horizon black hole metric in Einstein gravity after suitable coordinate transformations.

        \subsection{Asymptotic behaviour at \texorpdfstring{$\alpha\sim\mathcal{O}(1/n^2)$}{TEXT}}
    \label{subsec:ordless}

      \hspace{1cm}Let us attempt a coordinate transformation of the form used in Einstein gravity \cite{Ricciflow} on the conformal metric(\ref{conformal}) in \ref{formnearmergred}, obtained from (\ref{modifromegb}) and see what happens;
    \begin{equation}
        t=\frac{n\Tilde{t}}{2}
        \label{corst}
    \end{equation}
    \begin{equation}
        z=lnsinh(x/2)-\frac{n}{4}sin^2\theta
    \end{equation}
    \begin{equation}
        \rho=\frac{\sqrt{n}}{2}sin\theta\Big(1+\frac{2}{n}lncosh(x/2)\Big)
        \label{corend}
    \end{equation}

   Now, after these coordinate transformations, for a $V$ of the form (\ref{modifromegb}) we will get a metric of the form(at $\theta\sim O(1/\sqrt{n})$);

   \begin{multline}
        ds^2=\frac{n}{4}\Biggl[\Big(-tanh^2(x/2)\Big(1-\frac{\epsilon}{cosh^2(x/2)}\Big)+O(1/n)+O(\epsilon/n)\Big)d\Tilde{t}^2
        +\Big(\frac{1}{n^2}+O(\epsilon/n^2)+O(1/n^{2.5})\Big)dx^2\\
        +\Big(1+O(1/n)+O(\epsilon/n)\Big)d\Omega^2_{d-2}+\Big(O(1/n^{2.5})+O(\epsilon/n^{1.5})\Big)dxd\theta\Biggl]
    \label{blackholenearepsilon}
   \end{multline}

   \hspace{1cm}Now the above form of the metric is the near horizon limit of the EGB black hole \ref{blackholenear} with additional deformations\footnote{ This is up to overall scaling factors that can be absorbed in re-scaling of coordinates.} when we identify $\epsilon=\frac{\alpha n^2}{4r_h^2}$. Just as in the case of Einstein gravity\cite{Ricciflow}, the deformations can be attributed to the relevant geometry being that of a caged black hole instead of a black hole that asymptotically approaches Minkowski spacetime. As $\epsilon\longrightarrow0$ the deformations reduce to the Einstein gravity deformed black hole metric.

   \hspace{1cm}As the flow equation in EGB gravity is the same as in Einstein gravity up to  $\alpha\sim\mathcal{O}(1/n^2)$, the EGB black hole metric matches the near neck region asymptotically to leading order with $\epsilon=\frac{\alpha n^2}{4r_h^2}$ . This suggests that, like in Einstein gravity\cite{Ricciflow}, the phase transition from the black hole to the black string and vice versa is allowed. But there is a caveat here, which is that $\epsilon$ should be sufficiently small for the linearized approximation in the previous section to be valid.

   \subsection{Asymptotic behaviour at \texorpdfstring{$\mathcal{O}(1/n)>\alpha>\mathcal{O}(1/n^2)$}{TEXT}}
   \label{subsection:ordgat}
    \hspace{1cm}We had demonstrated earlier in section \ref{subsec:floeqbhbs} that if $V$ were a solution to the 'modified' logarithmic diffusion equation, it should either be of the form $V(\rho,z;\alpha)$ or it should be independent of $\rho$ and $\alpha$. Also, from the form of the metric in the conformal gauge (\ref{conformal}), it can be seen that $V$ is the term multiplying $dt^2$ under the coordinate transformation (\ref{corst}-\ref{corend}). But the term multiplying $dt^2$ in the near horizon limit of the EGB black hole is (\ref{formoff1}), which is not of the form $V(\rho,z;\alpha)$. It does not depend on $\alpha$ but depends on $\rho$. This means that a $V$ matching the EGB black hole metric asymptotically away from the near-neck region is not possible with a GB parameter $\mathcal{O}(1/n)>\alpha>\mathcal{O}(1/n^2)$. This would imply that the asymptotic geometry does not approach the geometry relevant for a black string-black hole phase transition. So there is a range of $\alpha$ in which the corrections are subordinate to the Einstein part, but a merger point like the one present in Einstein gravity where the geometry approaches a black hole away from the cone does not seem possible. We cannot rule out a topology changing phase transition, but we observe that the geometry does not approach that of the black hole. In the large dimension limit this occurs for $\mathcal{O}(1/n)>\alpha>\mathcal{O}(1/n^2)$.

 \subsection{Some speculations on copying of solutions}
    \label{subsec:nearcri}
\hspace{1cm}Notice that based on the observations made in sections \ref{subsec:ordless}\&\ref{subsection:ordgat}  we can see a range of $\alpha$, $\mathcal{O}(1/n)>\alpha>\mathcal{O}(1/n^2)$, for which while we may have a topology changing transition, however, the metric does not approach the black hole metric asymptotically. But for $\alpha \sim O(1/n^2)$, we can have a phase transition to the black hole but as demonstrated in section \ref{subsec:ordless}, the derivation of the asymptotic geometry requires the parameter $\epsilon$ to be small. The asymptotic geometry then approaches that of a black hole.

  \hspace{1cm}An important property related to black holes on cylinders in Einstein gravity is the copying of spherically symmetric static solutions to create new static solutions\cite{phasestructure}\cite{playstring}\cite{kblachhole}. This implies that one can copy the Einstein gravity merger point to create new merger points indexed by a natural number $k$. Copying is defined as following process of generating new solutions from known solutions;
        \begin{equation}
           d s^{2}=-f d t^{2}+\frac{L^{2}}{4\pi^2}[f^{-1} A d R^{2}+\frac{A}{K^{d-3}} d v^{2}+K R^{2} d \Omega_{d-3}^{2}],  \quad f=1-\frac{R_{0}^{d-4}}{R^{d-4}}
           \label{copyform}
       \end{equation}
   which gets mapped to a new solution by;
        \begin{equation}
            \tilde{A}(R, v)=A(k R, k v), \quad \tilde{K}(R, v)=K(k R, k v), \quad \tilde{R}_{0}=\frac{R_{0}}{k}
        \end{equation}
    ,for every natural number $k$,(the corresponding metric is obtained by substituting $\tilde{A},\tilde{K}\&\tilde{R}$ into (\ref{copyform}) for $A,K\&R$ respectively) under which the mass and tension scales as;
        \begin{equation}
            \tilde{M}=\frac{M}{k^{d-4}},   \quad\tilde{n}=n.
            \label{map}
        \end{equation}
    \hspace{1cm}The $k$ copied merger-point connects a $k$ black hole phase with a phase of non-uniform black strings through topology changing phase transitions where the horizon topology of a non-uniform black string ($S^1\times S^{d-3}$) changes to that of $k$ black holes each with a horizon topology of $S^{(d-2)}$\cite{kblachhole}.

    \hspace{1cm} In Einstein gravity, the existence of a single black hole merger point implies the existence of a $k$ black hole merger point through the copying of solutions. But is the same true if Gauss-Bonnet corrections are considered?

     \hspace{1cm}While the copying property has to be rigorously investigated in EGB gravity, let us assume such copying is possible. Then we shall attempt to express the condition that $\epsilon$ should be sufficiently small by trying to express it in terms of the three parameters then associated with the merger point and EGB gravity, which are the size of $S^1- L$, the copy number of the merger point $k$ and $\alpha$. We can do this by considering the $k$ copied non-uniform black string which we expect is connected through the merger point to a $k$ black holes solution. Since the neck of the merger point is in a region of extent $O(1/n)$ \cite{Ricciflow}, we have approximately;
    \begin{equation}
        L=k\Big[2r_h+O(\frac{1}{n})\Big]
    \end{equation}
   \hspace{1cm} where $r_h$ is the radius of one of the (identical) black holes formed just after the merger point. This condition implies that $\epsilon$ can be expressed as $\frac{k^2\alpha n^2}{L^2}$ in the large $n$ limit. Further, if we define $\alpha=Cn^{-2}$, then if $\frac{k^2 C}{L^2}$ is small, the geometry approaches the black hole geometry and we have a phase transition to a black hole. Then if copying of solutions is true in EGB gravity, we should also expect a merger geometry mediating a $k$ copied black string to a $k$ black hole phase transition.

   \hspace{1cm}It would be interesting to rigorously see if there is copying of solutions in EGB gravity and, if so, what would be the $k$'th copy of such a topology changing transition where the metric does not approach a black hole metric asymptotically. Further, how does the size $L$ of the $S^1$ affect the transition? Our perturbation parameter depending on $k\& L$ may indicate that we can expect novel features in the phase diagram for EGB gravity, but this must be investigated not just as a perturbation to the Einstein gravity solution, but non-perturbatively.

\section{Leading order Modifications to the Smooth Cone from EGB Gravity}
    \label{cone}
    \subsection{When \texorpdfstring{$\alpha\sim\mathcal{O}(1/n^2)$}{TEXT}}
    \label{cone1}
          \hspace{1cm}Let us attempt to find the leading order perturbative corrections in $\epsilon^{cone}$ to the large $n$ smooth cone by using the corresponding $V_0^{cone}=(\rho^2-\mu)/cosh^2z$, \cite{Ricciflow}. If we insist on spherical symmetry, this would require us to use an $F^{cone}$ of the form;
    \begin{equation}
        F^{cone}(\rho,z)=\frac{g(\rho)}{cosh^2z}
    \end{equation}

    \hspace{1cm}The above form of correction will ensure that the modifications will not have any angular dependence. We do not want angular dependence because in the euclidean version, we need a double cone over $S^2$\cite{phasetranseinpap}\cite{phasetransein}. This can be seen by doing the transformation $tan(\chi)=sinh(z)$, leading the 2-dimensional sub-space to reduce to the corresponding sphere in the euclideanized version maintaining the spherical symmetry.\footnote{The 2D subspace with conformally flat metric has been argued to have the topology of a sphere in the euclidean version of the metric\cite{phasetranseinpap}\cite{phasetransein}.}

    \hspace{1cm}Now substituting this form of $F^{cone}$ into (\ref{condpert}) with the corresponding $V_0^{cone}$ will result in $F^{cone}$ being zero, this means that the perturbation linear in $\epsilon$ is such that it is zero near the neck. This means that the Gauss Bonnet correction to the smooth cone is not present at linear order in $\epsilon$ for $\alpha\sim\mathcal{O}(1/n^2)$.

    \subsection{When \texorpdfstring{$\mathcal{O}(1/n)>\alpha>\mathcal{O}(1/n^2)$}{TEXT}}
    \label{cone2}
    \hspace{1cm} As we expect the Wick-rotated geometry of the two dimensional spacetime to be a sphere\cite{phasetranseinpap}\cite{phasetransein}, let us start with a metric of the form

    \begin{equation}
        V=\frac{(S_0+2n\alpha G(\rho))}{cosh^2(z)}
        \label{formconf}
    \end{equation}

    where $S_0=\rho^2-\mu$ \cite{Ricciflow}. Now the evolution equation(\ref{modlog}) up to leading order becomes

    \begin{equation}
        \partial_\lambda                 V=\partial^2_z(ln(V))+\frac{2n\alpha}{V_0^{cone}}\Big[(\partial_\lambda V_0^{cone})^2-3\partial_\lambda V_0^{cone}\partial^2_z(ln(V_0^{cone}))+(\partial^2_z(ln(V_0^{cone})))^2\Big]
    \label{modlogled}
    \end{equation}

    ,where $V_0^{cone}=\frac{S_0}{cosh^2z}$\cite{Ricciflow}, Now the above equation reduces to;

    \begin{equation}
        \partial_\lambda                 V=\partial^2_z(ln(V))-2n\alpha\times\frac{4sech^2(z)}{(\rho^2-\mu)}
    \label{modlogled2}
    \end{equation}
    Now upon plugging in $V$, (up to leading order)we get an equation of the form;
    \begin{equation}
        \partial_\rho G =\frac{4\rho}{(\rho^2-\mu)}
    \end{equation}

    This gives $V$ as;

    \begin{equation}
         V=\frac{(S_0+4n\alpha ln(\rho^2-\mu))}{cosh^2(z)}
         \label{lauranform}
    \end{equation}

    \hspace{1cm}So, in this range of $\mathcal{O}(1/n)>\alpha>\mathcal{O}(1/n^2)$, we have the leading order corrections to the cone to be of the above form. For $\rho$ close to $\mu$, the cone geometry receives large corrections due to the logarithmic factor and perturbation theory is no longer valid.

    \subsection{Conifold geometry in EGB gravity}
    \hspace{1cm}We will now consider solutions in EGB gravity of the form (\ref{dubconform}) and study the constraints imposed by EGB field equations on them. This is a cross-check on the computations in the previous section done with perturbation parameter $\epsilon$.
    \begin{equation}
        ds^2=N^2(\rho)d\rho^2+\frac{S_{GB}(\rho)}{n}d\Omega_2^2+\rho^2d\Omega_n^2
        \label{dubconform}
    \end{equation}
     For the above form of the metric, we can compute the following quantities up to leading order \footnote{We have only mentioned the explicit form of the leading order terms for quantities that are relevant to our analysis, for other quantities, we have only mentioned their order.} in a $1/n$ expansion;
     \begin{equation*}
         \begin{array}{l}
              R_{00}=n\frac{\partial_0ln(N)}{\rho};\quad R_{11}=\left(1-\frac{\dot{S}_{GB}}{2\rho N^2}\right)sin^2(\theta); \quad R_{22}=1-\frac{\dot{S}_{GB}}{2\rho N^2};\quad R_{ij}=n\left(1-\frac{1}{N^2}\right)g_{ij}\\
             \\
              \quad\quad\quad R=\frac{n^2}{\rho^2}\left(1-\frac{1}{N^2}\right);\quad\tensor{R}{^1_0_1_0}\sim O(1);\quad\tensor{R}{^2_0_2_0}\sim O(1);\quad\tensor{R}{^2_1_2_1}= sin^2(\theta)\\
             \\ \quad\quad\quad\quad\tensor{R}{_0_i_0^j}\sim O(1);\quad\tensor{R}{_1_i_1^j}\sim O(1/n);\quad\tensor{R}{_2_i_2^j}\sim O(1/n);\quad \tensor{R}{_i_j_k^l}\sim O(1)

         \end{array}
     \end{equation*}

     The EGB field equations are;

    \begin{equation}
        G_{\mu\nu}+\alpha Q_{\mu\nu}=0
    \end{equation}
    where;
\begin{equation}
    G_{\mu\nu}=R_{\mu\nu}-\frac{1}{2}g_{\mu\nu}R
\end{equation}
\begin{equation}
\begin{array}{r}
Q_{\mu \nu}=2\left[RR_{\mu \nu}-2 R_{\mu \alpha} \tensor{R}{_\nu^\alpha}-2 R^{\alpha \beta} R_{\mu \alpha \nu \beta}\right. \\
\left.+\tensor{R}{_\mu^\alpha ^\beta ^\gamma}R_{\nu \alpha \beta \gamma}\right]-\frac{1}{2} g_{\mu \nu} L_{EGB}
\end{array}
\end{equation}

   $ \bullet\quad\alpha\sim\mathcal{O}(1/n^2)$

     \hspace{1cm}For (\ref{dubconform}) the leading order term in $n$ in the large $n$ limit for $\alpha\sim\mathcal{O}(1/n^2)$, is always of the form $RR_{\mu\nu}$, and at leading order, the leading term for the Ricci scalar $R$ is zero when $N=1 +....$ where the ellipses refer to possible corrections at higher powers of $1/n$. Simply taking $N=1$ satisfies the leading order EGB field equations. The leading term in $S_{GB}$ in the large $n$ limit is not fixed at this order. For $\alpha \sim \mathcal{O}(1/n^2)$, we expect $S_{GB} = S_0 + O(1/n)$ and $S_0$ gets fixed by the next to linear order terms in the EGB field equations.

     \hspace{1cm}In this range of $\alpha$, we get at sub-leading order $R_{11}=0,R_{22}=0$, which will fix the leading part of $S_{GB} $ as $S_0 = \rho^2-\mu$. There are many nontrivial contributions at next-to next-to leading order in $1/n$, so it is not feasible to obtain the corrections to $N$ and $S_{GB}$ beyond leading order. We thus see that in this range of $\alpha$ we get the fused, critical and split cone geometry (for $\mu >0, \mu = 0$ and $\mu < 0$ respectively) obtained by Emparan and Suzuki \cite{Ricciflow} for Einstein gravity.

     $\bullet\quad\mathcal{O}(1/n)>\alpha>\mathcal{O}(1/n^2)$

      In this range, we will have the following equation in the large $n$ limit upon setting $N=1$ \footnote{Here the GB corrections are strictly speaking at order $n\alpha$ which is not $O(1)$, but since $n\alpha>1/n$, this is the next to leading order term and since this term is present because of the GB correction, we have written them together.};

     \begin{equation}
         \begin{array}{l}
              R_{11}-4\alpha\left(R_{11}\tensor{R}{^1_1}+R^{22}R_{1212}-\tensor{R}{_1^2^1^2}R_{1212}\right)=0\\
              R_{22}-4\alpha (R_{22}\tensor{R}{^2_2}+R^{11}R_{2121}-\tensor{R}{_2^1^2^1}R_{2121})=0
         \end{array}
     \end{equation}

     which can be written in terms of $S_{GB}$ as:

     \begin{equation}
\left(1-\frac{\dot{S}_{GB}}{2\rho}\right)-\frac{4 n\alpha}{S_{GB}}\left(\frac{(\dot{S}_{GB})^{2}}{4 \rho^{2}}-\frac{3}{2} \frac{\dot{S}_{GB}}{\rho}+1\right)=0
    \end{equation}

Unlike the previous range of $\alpha$, in this range, we have as a consequence of studying the order of various terms in the field equation,
\begin{equation}
     \begin{array}{l}
           S_{GB}=S_0+n\alpha S_1+O(1/n)\\
     \end{array}
     \end{equation}
 where the order of $n\alpha$ in the range of $\alpha$ given by $\mathcal{O}(1/n)>\alpha>\mathcal{O}(1/n^2)$ is greater than $O(1/n)$.

    \hspace{1cm} Now this equation will set $S_0=\rho^2-\mu$ and $S_1=4ln(\rho^2-\mu)$. But the perturbation expansion in $n$ is not valid when $\rho^2\sim\mu$. Thus this metric is not valid in the neck region. This divergence for $\rho^2\sim\mu$  is similar to the $O(1/n)$ corrections in Einstein gravity\cite{Ricciflow} where it is argued that a finer neck region solution is the relevant geometry sufficiently close to the neck, by doing the large $n$ limit in a slightly different way. In our case as well, that may be possible. However even if a topology changing transition does occur for this range of $\alpha$, the metric does not approach the EGB black hole metric asymptotically.

    \section{Conclusion}
    \label{sec:conclu}

     \hspace{1cm} In this paper, we have investigated the presence of a black hole black string phase transition via a merger point geometry in EGB gravity in the large dimension $n$ limit. We have performed an analysis of the near-neck geometry. It is conelike for $\alpha\sim \mathcal{O}(1/n^2)$ with the cone geometry being fused, critical or split cones derived in the context of Einstein gravity by Emparan and Suzuki \cite{Ricciflow}. Thus we certainly expect a topology changing phase transition in this case consistent with the arguments of Kol \cite{phasetranseinpap}-\cite{phasestructure}.
     For $\mathcal{O}(1/n)>\alpha>\mathcal{O}(1/n^2)$, our near-neck conical geometry receives a correction which become divergent closer to the neck. It is possible that this could be solved by going to a finer neck geometry as in \cite{Ricciflow} for Einstein gravity and a slightly different large $n$ limit. In the large dimension limit, we cannot match the asymptotic form of the near neck geometry to the EGB black hole metric when the Gauss-Bonnet parameter, $\mathcal{O}(1/n)>\alpha>\mathcal{O}(1/n^2)$ while it is possible to do so when $\alpha\sim \mathcal{O}(1/n^2)$. This suggests that for $\mathcal{O}(1/n)>\alpha>\mathcal{O}(1/n^2)$, while we cannot rule out a topology changing phase transition as predicted by Kol \cite{phasetranseinpap}-\cite{phasestructure}, a transition between the non-uniform black string geometry and the black hole geometry in EGB gravity does not seem to occur directly as the geometry away from the neck does not go over to the black hole metric. Thus it is an interesting question as to whether we can have a topology changing transition when the geometry does not approach the black hole metric and what would be the significance of such a solution. Is there an intermediate geometry after the merger point through which the transition to a black hole happens?

   \hspace{1cm} There are several extensions of this work that are possible. The rich phase space in the black hole black string phase transition remains to be explored analytically and numerically to understand the various branches of solutions in EGB gravity, as has been done in Einstein gravity. In particular, it would be interesting to see if there is the copying of solutions as in Einstein gravity. The dynamical evolution away from the Gregory-Laflamme instability in EGB gravity also needs to be investigated in the same manner as it was investigated in Einstein gravity in \cite{selfsimcosvil}.

\end{document}